\documentclass[aps,prb,twocolumn]{revtex4}
\usepackage{graphicx,amsmath}
\usepackage[english]{babel}

\begin{document}
\title{Pseudogap in the microwave response of YBa$_2$Cu$_3$O$_{7-x}$}
\author{M.R.~Trunin, Yu.A.~Nefyodov, and A.F.~Shevchun}
\affiliation{Institute of Solid State Physics RAS, 142432 Chernogolovka,
Moscow district, Russia}

\begin{abstract}
The in-plane and out-of-plane surface impedance and microwave
conductivity components of one and the same YBa$_2$Cu$_3$O$_{7-x}$
($0.07\le x\le 0.47$) single crystal are determined in the wide
ranges of temperature $T$ and carrier concentration $p$ in CuO$_2$
planes. The following features of the superfluid density
$n_s(T,p)\propto \lambda_{ab}^{-2}(T,p)$ are observed at $T<T_c/2$
and $0.078\le p\le 0.16$: (i) $n_s(0,p)$ depends linearly on $p$,
(ii) the derivative $|dn_s(T,p)/dT|_{T\to 0}$ depends on $p$
slightly in the optimally and moderately doped regions ($0.10<p\le
0.16$); however, it rapidly increases with $p$ further lowering
and (iii) the latter finding is accompanied by the linear
low-temperature dependence $\Delta n_s(T)\propto (-T)$ changing to
$\Delta n_s(T)\propto (-\sqrt{T})$. For optimum oxygen content the
temperature dependence of the normalized imaginary part of the
$c$-axis conductivity $\lambda_c^2(0)/\lambda_c^2(T)$ is found to
be strikingly similar to that of
$\lambda_{ab}^2(0)/\lambda_{ab}^2(T)$ and becomes more convex with
$p$ lowering. $\lambda_c^{-2}(0,p)$ values are roughly
proportional to the normal state conductivities $\sigma_c(T_c,p)$
along the $c$-axis. All these properties can be treated in the
framework of $d$-density wave order of pseudogap.

\end{abstract}
\maketitle

\section{Introduction}

Last years a lot of interest has been attracted by investigations
of the nature of pseudogap states of high-$T_c$ superconductors'
(HTSC) phase diagram. This area corresponds to lower concentration
$p$ of holes per copper atom in the CuO$_2$ plane and lower
critical temperatures $T_c$ in comparison with the optimal value
$p\approx 0.16$ and the maximum temperature $T_{c,max}$ of the
superconducting transition. The $p$ and $T_c$ values in HTSC
satisfy the following empirical relationship \cite{Tal1}:
$T_c=T_{c,max}[1-82.6(p-0.16)^2]$. Currently, the origin of the
pseudogap remains unclear. Proposed theoretical scenarios may be
divided into two categories. One is based on the idea that the
pseudogap is due to precursor superconductivity, in which pairing
takes place at the pseudogap transition temperature $T^*>T_c$ but
achieves coherence only at $T_c$. The other assumes that the
pseudogap state is not related to superconductivity per se, but
rather competes with it. This magnetic precursor scenario of the
pseudogap assumes dynamical fluctuations of some kind, such as
spin, charge or structural, or so-called staggered flux phase.
These two scenarios treat anomalies of electronic properties in
underdoped HTSC observed at temperatures both above $T_c$ and in
its vicinity~\cite{Timu,Tal2,Sado,Norm}.

In the heavily underdoped HTSC, at $T\ll T_c$ a competition of pseudogap
and superconducting order parameters develops most effectively and results
in the peculiarities of the superfluid density $n_s(T,p)$ as a function of
$T$ and $p$. It is well known that in clean BCS $d$-wave superconductors
(DSC) the dependence $\Delta n_s(T)\equiv n_s(T)-n_0$ is linear on
temperature $T\ll T_c$: $\Delta n_s(T)\propto (-T/\Delta_0)$, where
$n_0=n_s(0)$ and $\Delta_0=\Delta (0)$ are the superfluid density and the
superconducting gap amplitude at $T=0$. This dependence is confirmed by
the measurements of the $ab$-plane penetration depth
$\lambda_{ab}(T)=\sqrt{m^*/\mu_0e^2n_s(T)}$: $\Delta\lambda_{ab}(T)\propto
T$ at $T<T_c/3$, where $\mu_0$, $m^*$ and $e$ are the vacuum permeability,
the effective mass and the electronic charge, respectively. The derivative
$|\,dn_s(T)/dT|$ at $T\to 0$ determines $n_0/\Delta_0$ ratio. If thermally
excited fermionic quasiparticles are the only important excitations even
at $p<0.16$, then the slope of $n_s(T)$ curves at $T\ll T_c$ is
proportional to $n_0(p)/\Delta_0(p)$ ratio: $|\,dn_s(T)/dT|_{T\to
0}\propto n_0(p)/\Delta_0(p)$. The measurements of $\lambda_{ab}(0)$ in
underdoped HTSC showed that the superfluid density
$n_0(p)\propto\lambda_{ab}^{-2}(0)$ increases approximately linearly with
$p>0.08$ reaching its maximum value at $p\approx 0.16$~\cite{Lor,Ber}.

When decreasing $p<0.16$ and hence approaching the dielectric
phase, the role of electron correlations and phase fluctuations
becomes increasingly significant. The generalized Fermi-liquid
models (GFL) allow for this through $p$-dependent Landau parameter
$L(p)$~\cite{Lee,Mil1,Mil2} which includes $n_0(p)$. The values of
$\Delta_0(p)$ and $L(p)$ determine the doping dependence of the
derivative $|\,dn_s(T)/dT|_{T\to 0}=L(p)/\Delta_0(p)$. In
Ref.~\onlinecite{Lee} the ratio $L(p)/\Delta_0(p)$ does not depend
on $p$; the model~\cite{Mil1} predicts $L(p)/\Delta_0(p)\propto
p^{-2}$. The measurements of YBa$_2$Cu$_3$O$_{7-x}$ single
crystals~\cite{Bonn} and oriented powders~\cite{Pan} with the
holes concentration $p\gtrsim 0.1$ showed that the slope of
$n_s(T)$ dependences at $T\to 0$ is either slightly
$p$-dependent~\cite{Bonn}, which agrees with
Ref.~\onlinecite{Lee}, or diminishes with decreasing $p\leq
0.16$,~\cite{Pan} which contradicts the GFL
models~\cite{Lee,Mil1,Mil2}.
\begin{figure*}[!]
\centerline{\includegraphics[width=0.8\textwidth,clip]{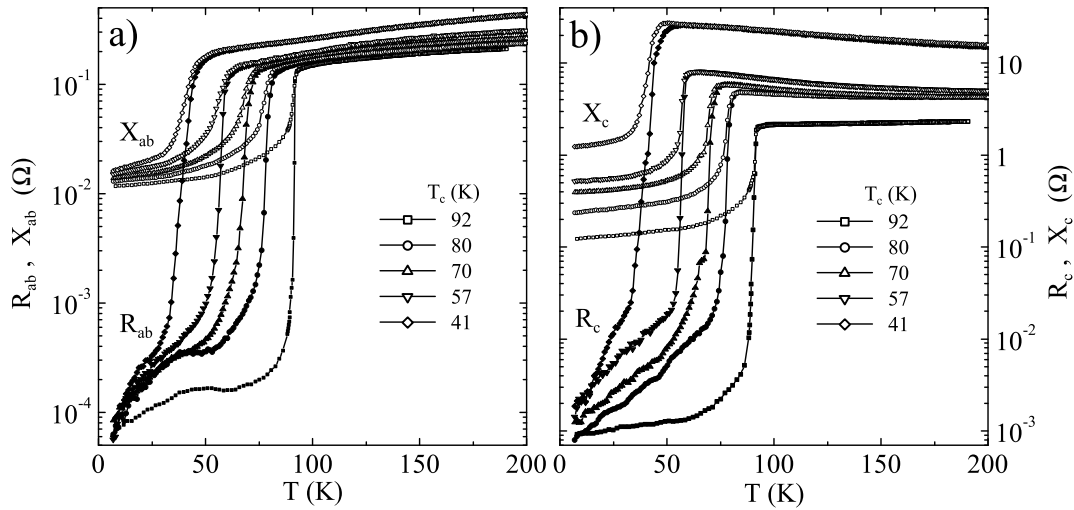}}
\caption{a) real $R_{ab}(T)$ (solid symbols) and imaginary
$X_{ab}(T)$ (open symbols) parts of the $ab$-plane surface
impedance of the five states of YBa$_2$Cu$_3$O$_{7-x}$ single
crystal; b) the components of the $c$-axis surface impedance.}
\label{f1}
\end{figure*}

In the precursor pairing model~\cite{Kost,Levi} of pseudogap, based on the
formation of pair electron excitations with finite momentum at $T^*>T_c$,
the influence of pseudogap order parameter on the quasiparticles spectrum
at $T<T_c$ leads to a rise of $\Delta_0(p)$ and decrease of $n_0(p)$ with
$p$ lowering. Hence, the decrease of the derivative $|\,dn_s(T)/dT|_{T\to
0}\propto n_0(p)/\Delta_0(p)$ is expected. The $n_s(T,p)/n_0$ dependences
calculated in Ref.~\onlinecite{Levi} show that their low temperature
slopes decrease with underdoping. An alternative behavior of
$|\,dn_s(T)/dT|$ follows from magnetic precursor $d$-density wave (DDW)
scenario of pseudogap~\cite{Chak}. In this model a DDW order parameter
$W(p,T)$ is directly introduced into the quasiparticle band structure. At
low energies the excitation spectrum of DDW consists of conventional
fermionic particles and holes like that of DSC with which it competes at
$p<0.2$. The DSC gap $\Delta_0(p)$ steadily vanishes with $p$ decreasing,
whereas the sum of zero-temperature squares $\Delta_0^2(p)+W_0^2(p)$
remains constant \cite{Tew}. In the issue, the DDW model predicts a growth
of the slope of $n_s(T,p)/n_0$ curves at low $T$ and $p<0.1$. At the same
time the opening of pseudogap influences weakly the $c$-axis penetration
depth $\lambda_c(T,p)$ of high-frequency field for currents running
perpendicular to CuO$_2$ planes. In particular, strong decrease of the
interlayer coupling integral $t_{\bot}(p)$ with $p$ lowering~\cite{Nyh} in
YBa$_2$Cu$_3$O$_{7-x}$ dominates over effects of the DDW order on
$\lambda_c(0,p)$~\cite{Kim}.

The present paper aims at the experimental verification of the above
mentioned theoretical speculations.

\begin{table}[b]
\vspace{-2mm}\caption{Annealing and critical temperatures, doping
parameters and penetration depths of YBa$_2$Cu$_3$O$_{7-x}$
crystal.}\vspace{-3mm}
{
\vspace{0.5cm}
\begin{tabular}{|c|c|c|c|c|c|c|}

\hline annealing&critical&\multicolumn{2}{c|}{doping}&
\multicolumn{2}{c|}{$\lambda$ values}&$\Delta\lambda_c(T)$\\
$T$, $^\circ$C&$T_c$, K&\multicolumn{2}{c|}{parameters}&
\multicolumn{2}{c|}{at $T=0$}&$\propto T^{\alpha}$\\
&&$p$&$x$&$\lambda_{ab}$, nm&$\lambda_{c}$, $\mu$m&$\alpha$\\

\cline{1-7}
500&92&0.16&0.07&152&1.55&1.0\\
\cline{1-7}
520&80&0.12&0.26&170&3.0&1.1\\
\cline{1-7}
550&70&0.106&0.33&178&5.2&1.2\\
\cline{1-7}
600&57&0.092&0.40&190&6.9&1.3\\
\cline{1-7}
720&41&0.078&0.47&198&16.3&1.8\\
\hline
\end{tabular}}\end{table}
\section{Experiment}

To fulfill the task, we investigated the anisotropy and evolution
of the temperature dependences of microwave conductivity
components in YBa$_2$Cu$_3$O$_{7-x}$ crystal under varying oxygen
doping in the range $0.07\le x\le 0.47$. The crystal of a
rectangular shape, with dimensions $1.6\times 0.4\times
0.1$~mm$^3$, has been grown in a BaZrO$_3$ crucible. The
measurements were made at the frequency of $\omega/2\pi=9.4$~GHz
and in the temperature range $5\le T\le 200$~K. To change an
oxygen content in the sample, we successively annealed the sample
in the air at different $T\ge 500^{\circ}$~C specified in Table~1.

According to susceptibility measurements at the frequency of
100~kHz, superconducting transition width amounted to 0.1~K in the
optimally doped state ($x=0.07$); however, the width increased
with the increase of $x$, having reached 4~K at $x=0.47$. The
temperatures of the superconducting transition were $T_c=92, 80,
70, 57, 41$~K which correspond to the concentrations $p=0.16,
0.12, 0.106, 0.092, 0.078$, respectively (Table~1). Anisotropy was
measured for each of the five crystal states. The whole cycle of
the microwave measurements included the following: (i) we measured
the temperature dependences of the quality factor and of the
frequency shift of the superconducting niobium resonator with the
sample inside in the two crystal orientations with respect to the
microwave magnetic field, transversal ($T$) and longitudinal
($L$); (ii) measurements in the $T$ orientation gave the surface
resistance $R_{ab}(T)$, reactance $X_{ab}(T)$ and conductivity
$\sigma_{ab}(T)=i\omega\mu_0/Z_{ab}^2(T)$ of the crystal cuprate
planes in its normal and superconducting states; (iii)
measurements in the $L$ orientation gave $\sigma_c(T)$, $X_c(T)$,
$R_c(T)$. See Ref.~\onlinecite{Nef1} for the details of the
measuring technique in the optimally doped YBa$_2$Cu$_3$O$_{6.95}$
crystal.

The temperature dependences of surface impedance
$Z_{ab}=R_{ab}+iX_{ab}$ components, $R_{ab}(T)$ and $X_{ab}(T)$,
are shown in Fig.~1a. In the normal state for each of the five
crystal states we have $R_{ab}(T)=X_{ab}(T)$ which implies the
validity of the normal skin-effect condition. The value of
residual losses $R_{ab}(T\rightarrow0)$ does not exceed
40~$\mu\Omega$. In the case of optimum oxygen content $R_{ab}(T)$
dependence has a broad peak at $T\sim T_c/2$ which vanishes with
$p$ lowering. At $T<T_c/3$ all $R_{ab}(T)$ curves are linear on
$T$. In Fig.~1b we demonstrate the temperature dependences of the
$c$-axis impedance components $R_c(T)$ and $X_c(T)$. The real and
imaginary parts of the surface impedance coincide at $T>T_c$,
$R_c(T)=X_c(T)$. Therefore, the resistivities $\rho_{ab}(T)$ and
$\rho_c(T)$ can be found from $R_{ab}(T)$ and $R_c(T)$ curves at
$T>T_c$ in Fig.~1, applying the standard formulas of the normal
skin effect: $\rho_{ab}(T)=2R_{ab}^2(T)/\omega \mu_0$,
$\rho_c(T)=2R_c^2(T)/\omega \mu_0$. Fig.~2 shows the evolution of
the dependences $\rho_{ab}(T)$ and $\rho_c(T)$ in
YBa$_2$Cu$_3$O$_{7-x}$ crystal with the change of $x$ in the
temperature range $T_c<T\le 200$~K.

\begin{figure}[t]
\centerline{\includegraphics[width=0.92\linewidth,clip]{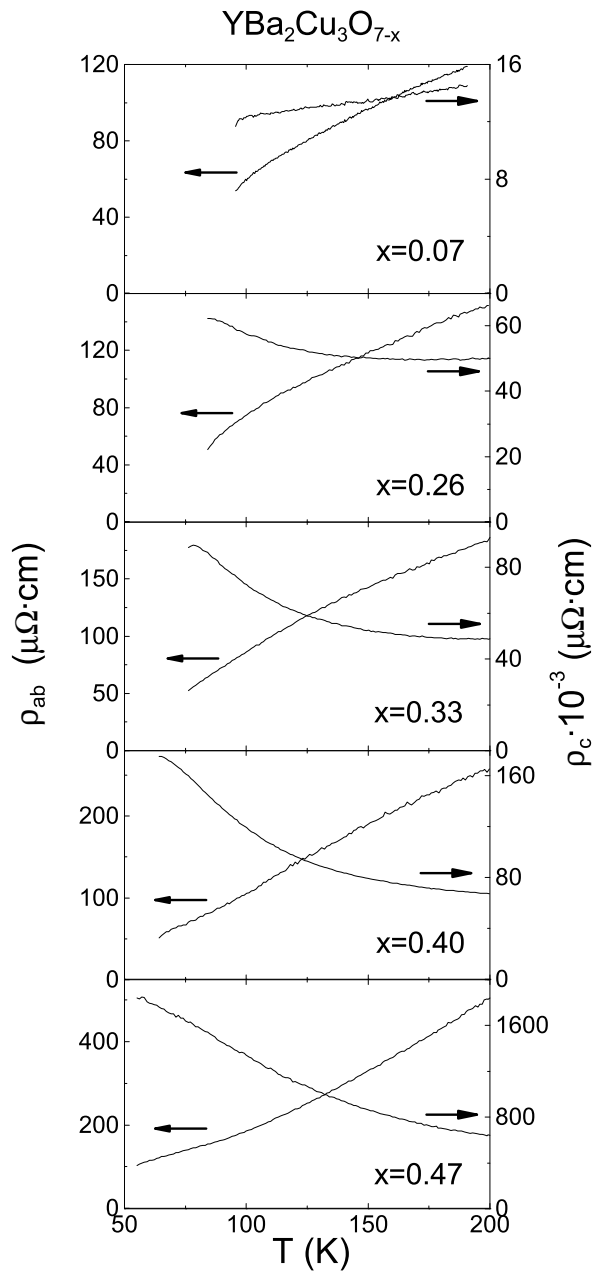}}\vspace{-3mm}
\caption{The evolution of the measured $\rho_{ab}(T)$ and
$\rho_c(T)$ dependences in YBa$_2$Cu$_3$O$_{7-x}$ with different
oxygen content.} \label{f2}
\end{figure}

\section{Results and discussion}
\begin{figure}[t]
\centerline{\includegraphics[width=0.9\linewidth,clip]{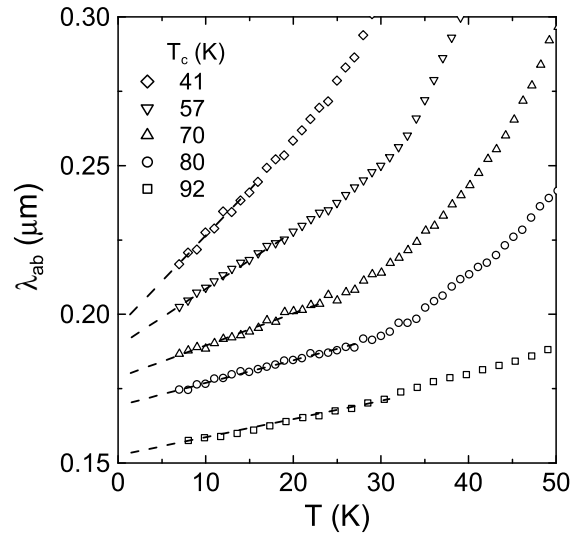}}
\caption{Low-temperature dependences of $\lambda_{ab}(T)$ (open
symbols) measured for five states of YBa$_2$Cu$_3$O$_{7-x}$
crystal with $T_c=92$~K, $T_c=80$~K, $T_c=70$~K, $T_c=57$~K, and
$T_c=41$~K. Dashed lines are linear extrapolations at $T<T_c/3$.}
\label{f3}
\end{figure}
Fig.~3 shows the low temperature sections of the measured
$\lambda_{ab}(T)=X_{ab}(T)/\omega\mu_0$ curves. The linear
extrapolation (dashed lines) of these dependences at $T<T_c/3$
gives the following $\lambda_{ab}(0)$ values: 152, 170, 178, 190,
198~nm for $p=0.16, 0.12, 0.106, 0.092, 0.078$, respectively
(Table~1). The error in $\lambda_{ab}(T)$ is largely determined by
the measurement accuracy of the additive constant $X_0$ which is
equal to the difference between the measured reactance shift
$\Delta X_{ab}(T)$ and $R_{ab}(T)$ at $T>T_c$~\cite{Tru03}.
\begin{figure}[t]
\centerline{\includegraphics[width=0.9\linewidth,clip]{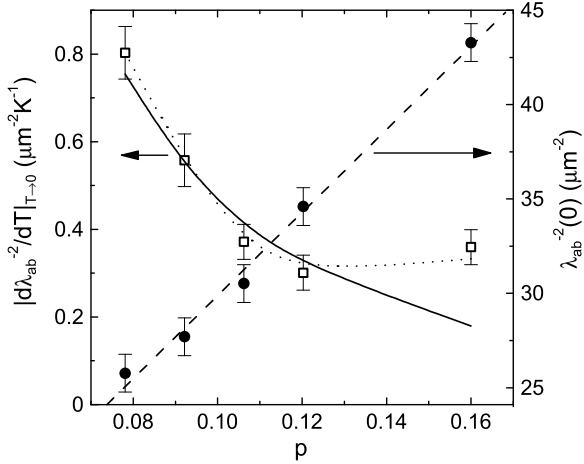}}
\caption{The values of $\lambda_{ab}^{-2}(0)=n_s(0)\mu_0e^2/m^*$
(right scale) and slopes $|\,d\lambda_{ab}^{-2}(T)/dT|_{T\to
0}=\mu_0e^2/m^*|\,dn_s(T)/dT|_{T\to 0}$ (left scale) as a function
of doping $p=0.16-\sqrt{(1-T_c/T_{c,max})/82.6}$ with
$T_{c,max}=92$~K in YBa$_2$Cu$_3$O$_{7-x}$. Error bars correspond
to experimental accuracy. The dashed and dotted lines guide the
eye. The solid line is $|\,dn_s(T)/dT|\propto p^{-2}$ dependence.}
\label{f4}
\end{figure}
\begin{figure}[b]
\centerline{\includegraphics[width=0.9\linewidth,clip]{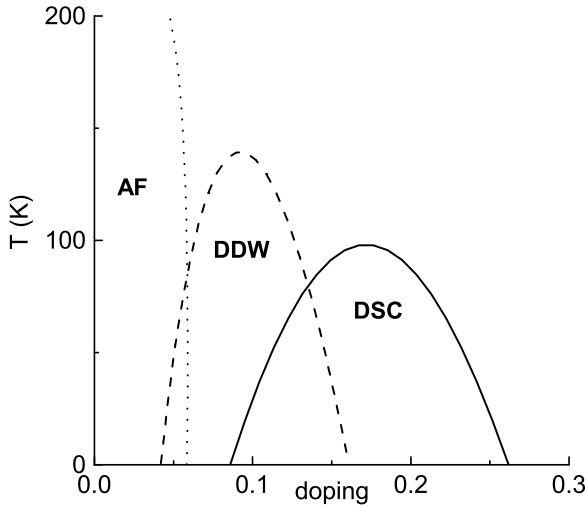}}
\caption{The temperature versus doping $p$ schematic phase diagram
based on calculations of Ref.~\cite{Nay}. AF is the
three-dimensional antiferromagnetic phase. The system is an
isolator in the AF state, a metal in the DDW and DDW+AF states,
and a superconductor in the DSC and DDW+DSC states \cite{Nay}.}
\label{f5}
\end{figure}

\begin{figure}[t]
\centerline{\includegraphics[width=0.9\linewidth,clip]{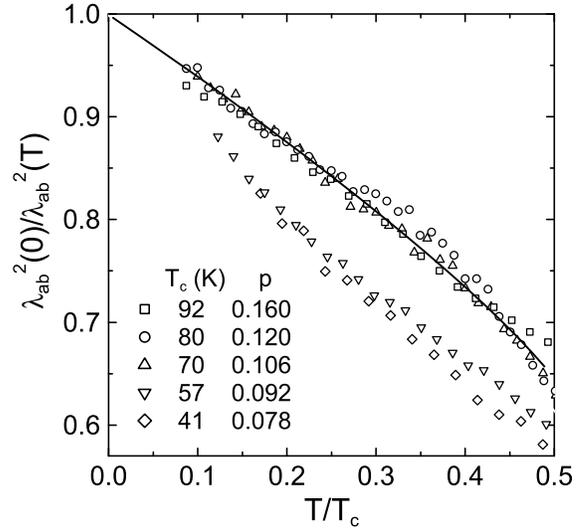}}
\caption{The measured dependences of
$\lambda_{ab}^2(0)/\lambda_{ab}^2(T)=n_s(T)/n_s(0)$ at $T<T_c/2$
in YBa$_2$Cu$_3$O$_{7-x}$ with different doping. The solid line is
the $\lambda_{ab}^2(0)/\lambda_{ab}^2(T)$ dependence in BCS
$d$-wave superconductor (DSC).} \label{f6}
\end{figure}

As follows from Fig.~4, halving of the concentration (namely, from
$p=0.16$ to $p=0.078$) results in approximately two times smaller
$\lambda_{ab}^{-2}(0)=n_0\mu_0e^2/m^*$ value. Similar behavior
$n_0(p)\propto p$ within the range $0.08<p\le 0.16$ was observed by other
groups~\cite{Lor,Ber}. It is easily seen that this dependence contradicts
Uemura's relation $n_0(p)\propto T_c(p)$ \cite{Uem}. The naive linear
extrapolation of the dashed line in Fig.~4 at $p<0.08$ leads to
nonphysical result: $n_0(p)$ is finite at vanishing $p$. To the best of
our knowledge there is no data of superfluid density measurements in HTSC
at $p<0.08$. As for theoretical predictions, $n_0$ linearity on $p$
extends down to $p=0$ in the model~\cite{Lee}, while in the DDW
scenario~\cite{Tew,Wan} it exists in the underdoped range of phase diagram
(Fig.~5) where the DSC order parameter grows from zero to its maximal
value, moreover, $n_0(p)$ is nonzero as $\Delta_0(p)$ vanishes (Fig.~1
from Ref.~\cite{Wan}). The latter agrees with our data.

In Fig.~4 we also show the slopes $|\,d\lambda_{ab}^{-2}(T)/dT|_{T\to
0}\propto |\,dn_s(T)/dT|_{T\to 0}$ of $\lambda_{ab}^{-2}(T)$ curves
obtained from $\lambda_{ab}(T)$ dependences at $T<T_c/3$. The value of
$|\,d\lambda_{ab}^{-2}(T)/dT|$ changes slightly at $0.1<p\leq 0.16$ in
accordance with Ref.~\cite{Lee}. However, it grows drastically at
$p\lesssim 0.1$, namely, $\lambda_{ab}^{-2}(T)$ slope increases 2.5 times
with $p$ decrease from 0.12 to 0.08. $|\,d\lambda_{ab}^{-2}(T)/dT|\propto
p^{-2}$ dependence~\cite{Mil1} is shown by solid line in Fig.~4 and
roughly fits the data at $p\leq 0.12$. The dotted line drawn through
$|\,d\lambda_{ab}^{-2}(T)/dT|$ experimental points in Fig.~4 qualitatively
agrees with the behavior of this quantity in the DDW model~\cite{Tew,Wan}.

The temperature dependence of the superfluid density $n_s(T)$ at low $T$
in the heavily underdoped YBa$_2$Cu$_3$O$_{7-x}$ proves to be one more
check-up of the DDW scenario of pseudogap.
$\lambda_{ab}^2(0)/\lambda_{ab}^2(T)=n_s(T)/n_0$ dependences obtained from
the data in Fig.~3 are shown in Fig.~6 for different $p$ values. The solid
line represents the DSC result. The evident peculiarities in Fig.~6 are
the concavity of $n_s(T)/n_0$ curves corresponding to the heavily
underdoped states ($p=0.078$ and $p=0.092$) and their deviation from DSC
and the curves for $p=0.16, 0.12, 0.106$. This behavior of the superfluid
density $n_s(T)/n_0$ contradicts the conclusions of the precursor pairing
model~\cite{Levi}, but agrees with the DDW scenario~\cite{Tew}. According
to~\cite{Tew}, at temperatures much smaller than the relevant energy
scales $W_0$ and $\Delta_0$, only the nodal regions close to the points
($\pi/2,\pi/2$) and symmetry-related points on the Fermi surface will
contribute to the suppression of the superfluid density. In a wide range
of temperatures $n_s(T)$ dependence will be linear for the optimally and
moderately doped samples, in which $\Delta_0$ is larger than or comparable
to $W_0$ (Fig.~5) and plays a leading role in the temperature dependence
of the superfluid density. However, for the heavily underdoped samples the
situation is quite different. As the DDW gap is much larger than the
superconducting gap in these heavily underdoped samples, $W_0$ becomes
dominant around the nodes. Though in the asymptotically low-temperature
regime the suppression of the superfluid density is linear on temperature,
there is an intermediate temperature range over which the suppression
actually behaves as $\sqrt T$. It is worth emphasizing that the authors of
Ref.~\cite{Tew} state that these features are independent of the precise
$W_0(p)$ and $\Delta_0(p)$ functional forms. The only input that is needed
is the existence of DDW order which diminishes with $p$ increase and
complementary development of the DSC order. The DDW order eats away part
of the superfluid density from an otherwise pure DSC system. Actually, in
the intermediate temperature range $0.1\,T_c<T\lesssim 0.5\,T_c$ the
experimental $n_s(T)$ curves in YBa$_2$Cu$_3$O$_{6.60}$ and
YBa$_2$Cu$_3$O$_{6.53}$ with $p<0.1$ are not linear but similar to
$\sqrt{T}$-dependences. This is confirmed by Fig.~7, where the measured
curves $\lambda_{ab}^{-2}(T)\propto n_s(T)$ are compared with linear
($\propto T$) in YBa$_2$Cu$_3$O$_{6.67}$ ($p=0.106$) and
$\sqrt{T}$-dependences $\Delta\lambda_{ab}^{-2}(T)=-3\sqrt{T}$
($\lambda_{ab}$ and $T$ are expressed in $\mu$m and K) in
YBa$_2$Cu$_3$O$_{6.60}$ ($p=0.092$) and
$\Delta\lambda_{ab}^{-2}(T)=-3.5\sqrt{T}$ in YBa$_2$Cu$_3$O$_{6.53}$
($p=0.078$). Dashed lines in Fig.~7 correspond to the linear at $T<T_c/3$
dependences of $\lambda_{ab}(T)$ presented in Fig.~3 and extended to
higher temperatures.
\begin{figure}[t]
\centerline{\includegraphics[width=0.94\linewidth,clip]{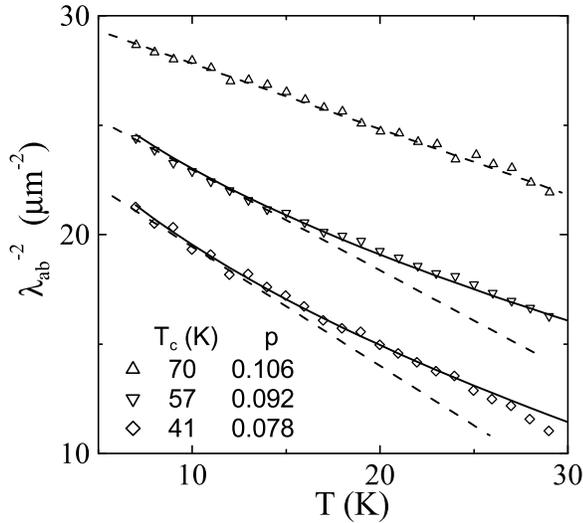}}
\caption{Comparison of experimental $\lambda_{ab}^{-2}(T)\propto n_s(T)$
curves (symbols) with linear $\Delta\lambda_{ab}^{-2}(T)\propto (-T)$
(dashed lines) and root $\Delta\lambda_{ab}^{-2}(T)\propto (-\sqrt T)$
(solid lines) dependences for moderately doped ($p=0.106$, $x=0.33$) and
heavily underdoped ($p=0.092$, $x=0.40$; $p=0.078$, $x=0.47$)
YBa$_2$Cu$_3$O$_{7-x}$.} \label{f7}
\end{figure}

\begin{figure}[t]
\centerline{\includegraphics[width=0.89\linewidth,clip]{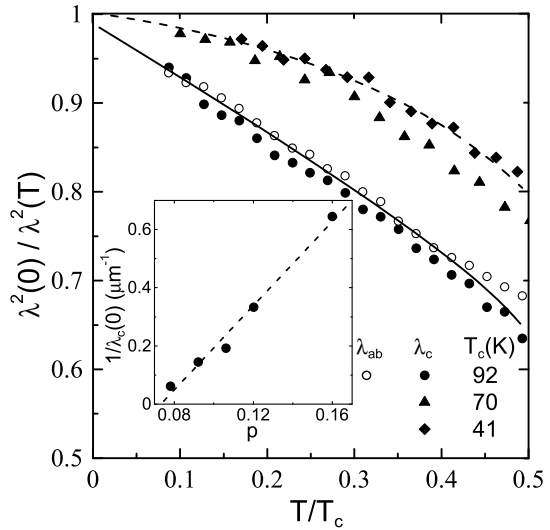}}\vspace{-3mm}
\caption{The dependence $\lambda_{ab}^2(0)/\lambda_{ab}^2(T)$
(open symbols) in YBa$_2$Cu$_3$O$_{6.93}$ and
$\lambda_c^2(0)/\lambda_c^2(T)$ (full symbols) measured for three
states of the YBa$_2$Cu$_3$O$_{7-x}$ crystal with $T_c=92$~K,
$T_c=70$~K and $T_c=41$~K. Solid and dashed lines stand for the
dependences $\lambda_c^2(0)/\lambda_c^2(T)$ calculated in
Ref.~\onlinecite{Lev1} for YBa$_2$Cu$_3$O$_{7-x}$ with different
oxygen deficiency. The inset shows $1/\lambda_c$ at $T=0$ as a
function of doping $p$.} \label{f8}
\end{figure}

It is also interesting to note that these deviations of
$\Delta\lambda_{ab}^{-2}(T)$ in YBa$_2$Cu$_3$O$_{6.60}$ and
YBa$_2$Cu$_3$O$_{6.53}$ are accompanied by inflection of the
resistivity $\rho_{ab}(T)$ curves in the normal state of these
samples. These inflections are seen at two lower $\rho_{ab}(T)$
curves in Fig.~2 around $T\sim 100$~K. One more feature of the
curves in Fig.~2 is that only the optimally doped
YBa$_2$Cu$_3$O$_{6.93}$ shows that both dependences $\rho_{ab}(T)$
and $\rho_c(T)$ have a metallic behavior, and the ratio
$\rho_c/\rho_{ab}$ approaches the anisotropy of effective masses
of charge carriers $m_c/m_{ab}=\lambda_c^2(0)/\lambda_{ab}^2(0)$
in $3D$ London superconductor, which type YBa$_2$Cu$_3$O$_{6.93}$
belongs to. The other states of YBa$_2$Cu$_3$O$_{7-x}$ have the
resistivity $\rho_c(T)$ increase with the decrease of temperature,
so that decrease of carriers concentration in
YBa$_2$Cu$_3$O$_{7-x}$ crystal results in the crossover from Drude
$c$-axis conductivity to the tunneling one~\cite{Tru2}. The
evolution of the temperature dependences of $\rho_c(T)$ with
doping correlates with those of the $c$-axis penetration depth
$\lambda_c(T)$. Solid symbols of Fig.~8 show the dependences
$\lambda_c^2(0)/\lambda_c^2(T)$ at $T\le T_c/2$ for
YBa$_2$Cu$_3$O$_{7-x}$ states with $T_c=92$~K, $T_c=70$~K and
$T_c=41$~K. Table~1 contains the values of the penetration depth
$\lambda_c(0)$ at $T=0$ and the exponents $\alpha$ in the measured
$\lambda_c(T)-\lambda_c(0)=\Delta\lambda_c(T)\propto T^\alpha$
dependences at $T\le T_c/3$. The peculiarity of the optimally
doped state YBa$_2$Cu$_3$O$_{6.93}$ is good coincidence of
$\lambda_{ab}^2(0)/\lambda_{ab}^2(T)$ (open circles in Fig.~8) and
$\lambda_c^2(0)/\lambda_c^2(T)$ temperature dependences. With the
decrease of $p$ the temperature dependence of
$\lambda_c^2(0)/\lambda_c^2(T)$ becomes substantially weaker than
that of $\lambda_{ab}^2(0)/\lambda_{ab}^2(T)$. Model~\cite{Lev1}
associates the reduction in the low-$T$ slope of
$\lambda_c^2(0)/\lambda_c^2(T)$ curves and the appearance of
semiconducting-like temperature dependence of $\rho_c(T)$ with a
decrease of the interlayer coupling in the crystal. Dashed line in
Fig.~8 represents numerical result \cite{Lev1} for this case. On
the other hand, in the optimally doped YBa$_2$Cu$_3$O$_{6.93}$ the
interlayer coupling is strong and quasiparticle transport along
the $c$-axis becomes identical to one in the anisotropic $3D$
superconductor \cite{Xiang}. Solid line in Fig.~8 is
$\lambda_c^2(0)/\lambda_c^2(T)$ dependence, calculated in
Ref.~\cite{Lev1} for this particular case. So, the low-$T$
dependences of $\lambda_c(T)$ are well described without taking
pseudogap effects into consideration. Let us consider now their
possible manifestations in the doping dependence of the $c$-axis
penetration depth.

From the inset to Fig.~8 follows that reciprocal value of the
zero-temperature penetration depth $1/\lambda_c(0,p)$ is roughly
linear on $p$. Note that it vanishes at $p\approx 0.07$ which is
near the value where $T_c$ does too (Fig.~5). There are several
theoretical models~\cite{Lev1,Hir} and experimental
confirmations~\cite{Bas} of the direct proportionality of
$\lambda_c^{-2}(0)$ to the $c$-axis conductivity $\sigma_c(T_c)$
in HTSC. In the simplest theory this correlation is caused by
$\lambda_c^{-2}\propto J_c$ relation, where $J_c$ is the $c$-axis
critical current in the $d$-wave superconductor with anisotropic
interlayer scattering and weak interlayer coupling. The value of
$J_c(0)$ is determined by both the superconducting gap $\Delta_0$
and the conductivity $\sigma_c(T_c)$. The symbols in the lower
inset to Fig.~9 show our data fitted by the dashed line
$\log\lambda_c(0)[\mu m]
=-0.48\log\sigma_c(T_c)[\Omega^{-1}m^{-1}]+2.08$. The latter
constant defines a proportionality factor $U_0(p)$ in
$\lambda_c^{-2}(0,p)=U_0(p)\,\sigma_c(T_c,p)$ relation. In the
framework of DDW model the value of $U_0(p)$ is determined by the
doping dependences of $\Delta_0(p)$, $W_0(p)$ and chemical
potential $\mu(p)$. As it is shown in Ref.~\cite{Kim} the opening
of DDW gap can lead to increase as well as to decrease of
$U_0(p)$. This depends on the position of Fermi surface with
respect to DDW gap, but in any case $U_0(p)$ changes less than
twice in the whole range of doping. The values of
$\lambda_c^2(0.16)/\lambda_c^2(p)$ at $T=0$ and
$\sigma_c(p)/\sigma_c(0.16)$ at $T=T_c$ are shown in Fig.~9. Their
ratio $U_0(p)/U_0(0.16)$ is demonstrated in the upper unset to
Fig.~9. This weak doping dependence indicates that the
contribution of the interlayer coupling integral
$t_{\bot}(p)\propto\sigma_c(T_c,p)$ into $\lambda_c(0,p)$ is
dominant.
\begin{figure}[t]
\centerline{\includegraphics[width=0.9\linewidth,clip]{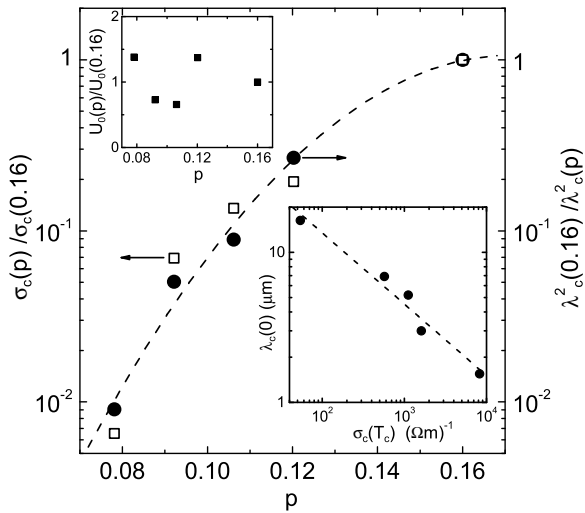}}\vspace{-3mm}
\caption{Doping dependences of $\lambda_c^2(p)/\lambda_c^2(0.16)$
at $T=0$ and $\sigma_c(p)/\sigma_c(0.16)$ at $T=T_c$. Their ratio
$U_0(p)/U_0(0.16)$ is shown in the upper inset. The lower inset is
$\lambda_c(0)$ versus $\sigma_c(T_c)$ plot.}\label{f9}
\end{figure}

Thus, four main experimental observations of this paper, viz, (i) linear
dependence of $n_0(p)$ in the range $0.078\le p\le 0.16$, (ii) drastic
increase of low-temperature $n_s(T)$ slope at $p<0.1$, (iii) the deviation
of $\Delta n_s(T)$ dependence from universal BCS behavior $\Delta
n_s(T)\propto (-T)$ at $T<T_c/2$ towards $\Delta n_s(T)\propto
(-\sqrt{T})$ with decreasing $p<0.1$, and (iv) very weak influence of
pseudogap on the low-$T$ and doping dependences of the $c$-axis
penetration depth evidence the DDW scenario of electronic processes in
underdoped HTSC. Nevertheless, the measurements of $\lambda_{ab}(T)$ and
$\lambda_c(T)$ at lower temperatures and in the high-quality samples with
smaller carrier density are necessary for ultimate conclusion.

Helpful discussions with A.I.~Larkin, E.G.~Maksimov, and Sudip Chakravarty
are gratefully acknowledged. This research was supported by RFBR grants
Nos. 03-02-16812 and 02-02-08004.

\end{document}